# New method for evaluating fitness using the waist-to-height ratio among Korean adults

Nam. Lyong Kang*

Department of Nanomechatronics Engineering, Pusan National University, Miryang 50463, Republic of Korea

## Abstract

**Objectives:** This paper introduces a new method for evaluating fitness and determining effective exercises for reducing abdominal obesity in Korean adults using the new kind of waist-to-height ratio (WHT2R).

**Materials and Methods:** The body mass index (BMI), body shape index (ABSI), and two other waist-to-height ratios (WHT.5R, WHTR) were considered as possible contenders for the WHT2R. The correlation coefficients were calculated by correlation analyses between the indices and four fitness tests for comparison. The LMV (lump mean value) and FSPW (fitness sensitivity percentage to WHT2R) were introduced to find the association between fitness and abdominal obesity using a linear regression method and to use as an indicator for the effective control of abdominal obesity.

**Results:** The WHT2R is more suitable for assessing fitness than the other indices and can be controlled effectively by decreasing the 10-m shuttle run score for both males and females.

**Conclusions:** The WHT2R can be used as a possible contender for evaluating fitness and is an effective indicator for the reduction of abdominal obesity. The LMV and FSPW can be used to establish personal exercise aims.

# 1 INTRODUCTION

Fitness is important because physical activity or exercise can reduce the level of abdominal obesity associated with cardiovascular diseases, such as hypertension (Carr et al., 2004; Dubbert et al., 2002; Lakka et al., 1994; Matsuzawa, 2010; Zhu, 2011). Lavie et al. (2019) examined the adverse effects of physical inactivity on the prevention of cardiovascular diseases in a study on the pathophysiological effects of weight gain on the cardiovascular system. Therefore, an effective index is necessary for a precise assessment of fitness and as an effective means of preventing and treating abdominal obesity. Hydrostatic densitometry is considered a good tool for assessing obesity, but it is inconvenient. More convenient and less expensive anthropometric indices, such as the body mass index (BMI) and waist-to-height ratios, have been used as alternative indicators of obesity and cardiovascular diseases.

The BMI is internationally accepted as a standard for assessing obesity (Hu et al., 2004; Kawai et al., 2019; Pasco et al., 2014; Raustorp et al., 2004). The advantages of the BMI are that it is not intrusive and easy to calculate with acceptable accuracy. On the other hand, abdominal obesity may be tied more closely to metabolic risks than the BMI (Blair et al., 1984; Kaplan, 1989; Hsieh et al., 1995; 1990; Lee et al., 1995; Smalley et al., 1990). Therefore, many studies have considered the waist-to-height ratio (WHtR), which is defined as the waist circumference (WC) divided by the height, as an indicator of abdominal obesity or a parameter to predict the risk of cardiovascular diseases (Choi et al., 2018; Hsieh et al., 2003; Shen et al., 2017; Vasquez et al., 2019). Nevill et al. (2017) introduced a new waist-to height ratio (WHT.5R) as the WC divided by the square root of the height and found it to be the best anthropometric index of the cardiometabolic risk. Bustamante et al. (2015) suggested that the reciprocal Ponderal index is a more appropriate index to assess fitness. Krakauer et al. (2012) developed a body shape index (ABSI), which is defined as $WC/(BMI^{2/3} \times height^{1/2})$, based on the WC adjusted for height and weight, and used it as a substantial risk factor for premature mortality.

Many researchers have considered the BMI, ABSI, WHT.5R, and WHTR as indices for measuring abdominal obesity and predicting the risk of cardiovascular diseases (Hsieh et al., 2003; Nevill et al., 2017; Zhu et al., 2002). Among these, the WHT.5R and WHtR are used widely as useful indices for predicting the risk of cardiovascular diseases because the BMI cannot represent the central fat distribution, and the waist circumference cannot differentiate between the visceral fat and subcutaneous fat (Ashwellet et al., 2012; Lu et al., 2006; Taylor et al., 1998).

This paper presents a new waist-to height ratio for assessing fitness, WHT2R, which is defined as the WC divided by the square of the height. The WC is used instead of weight in the BMI to indicate abdominal obesity. The correlation coefficients between the different indices and fitness tests were calculated to determine which among the BMI, ABSI, and three waist-to-height ratios (WHT.5R, WHtR, and WHT2R) has the best association with fitness. The association between WHT2R and fitness was also investigated per generation. On the other hand, it was impossible to find the dependence of fitness on the index individually because the distribution of fitness with respect to the index is quite complicated. Therefore, this study used the lump mean value (LMV), which was introduced by the present author, to determine the BMI for best fitness (Hwang et al., 2020). The WHT2R-dependence of the fitness test scores was investigated using the LMVs and a linear regression method, and the results were then compared with the BMI-dependence of the fitness test scores. The fitness sensitivity percentage to WHT2R (FSPW) was introduced to examine the sensitivities of fitness to the WHT2R and to determine the useful exercise for the effective control of abdominal obesity.

## 2  MATERIALS AND METHODS

### 2.1  Study design and participants

This cross-sectional study was performed using the public-use releases of the 2017 survey of national physical fitness (ISBN 979-11-952035-6-7) by the Korea Institute of Sport Science of the Korea Sports Promotion Foundation (KSPO). The survey is a nationally representative cross-sectional survey (http://www.sports.re.kr, available in Korean). Statistics Korea (National Statistics No 113004) approved the collection of data and samples from the cohorts participating in the study. All research participants provided written consent to take part.

The fitness tests considered were sit-up, standing long jump (SLJ), 20-m multi-stage shuttle run (20-m MSSR), and 10-m shuttle run (10-m SR) tests. The sit-up, SLJ, and 20-m MSSR tests measure the level of muscular endurance, speed, and cardiorespiratory endurance, respectively. The 10-m SR test measures the speed and agility (see Hwang et al., 2020 for details). The final study cohort comprised of 2082 males and 2038 females from a total of 4296 subjects aged 19 to 64 years. The subjects were excluded if they had extreme values of the five indices or four fitness tests. Table 1 lists the ranges considered.

**TABLE 1** Ranges of BMI, ABSI, WHT.5R, WHtR, WHT2R, Sit-up, SLJ, 20-m MSSR, and 10-m SR considered.

| Quantity [unit] | Men | | Women | |
|---|---|---|---|---|
| | Min. | Max. | Min. | Max. |
| BMI [kg/m$^2$] | 18.0 | 35.7 | 16.0 | 34.9 |
| ABSI [cm$^{11/6}$/kg$^{2/3}$] | 0.58 | 0.96 | 0.56 | 1.02 |
| WHT.5R [cm$^{0.5}$] | 5.03 | 9.17 | 4.57 | 8.62 |
| WHTR $\times 10^{-2}$ [none] | 38.5 | 69.3 | 36.1 | 67.6 |
| WHT2R $\times 10^{-4}$ [cm$^{-1}$] | 22.0 | 39.9 | 21.1 | 43.6 |
| Sit-up [times/minute] | 2 | 72 | 2 | 65 |
| SLJ [cm] | 61 | 280 | 50 | 223 |
| 20-m MSSR [times] | 3 | 100 | 2 | 82 |
| 10-m SR [sec] | 8.5 | 19.3 | 9.5 | 21.4 |

† Abbreviations: BMI = body mass index; ABSI = a body shape index; WHT.5R = waist circumference divided by the square root of the height; WHtR = waist-to-height ratio; WHT2R = waist circumference divided by the square of the height; SLJ = standing long jump; 20-m MSSR = 20-m multi-stage shuttle run; 10-m SR = 10-m shuttle run.

## 2.2 Analysis

This study examined which among the BMI, ABSI, WHT.2R, WHTR, and WHT2R has the best association with the fitness tests and was carried based on many studies, where the WC was more associated with the cardiometabolic mortality than the BMI, and height had an inverse association with the mortality (Langenberg et al., 2005; Petursson et al., 2011; Seidell et al., 2010). The WHT2R is defined using the WC instead of weight in BMI to consider the abdominal obesity as follows:

$$\text{WHT2R} \equiv \frac{\text{waist circumference}}{(\text{height})^2} \tag{1}$$

**TABLE 2** Lump mean values of the WHT2R ($\times\ 10^{-4}$ cm$^{-1}$), Sit-up (times/minute), SLJ (cm), 20-m MSSR (times), and 10-m SR (second) for males.

| N | WHT2R | | | Sit-up | | SLJ | | 20-m MSSR | | 10-m SR | |
|---|---|---|---|---|---|---|---|---|---|---|---|
| | Range | Mean | SD | Mean | SD | Mean | SD | Mean | SD | Mean | SD |
| 60 | 21.98-23.49 | 22.93 | 0.43 | 45 | 11.5 | 218 | 26.8 | 47 | 19.3 | 11.0 | 1.36 |
| 78 | 23.53-24.49 | 24.08 | 0.28 | 43 | 13.1 | 218 | 28.0 | 48 | 20.2 | 10.9 | 1.34 |
| 81 | 24.50-24.98 | 24.75 | 0.14 | 43 | 12.3 | 212 | 23.3 | 47 | 20.1 | 11.0 | 1.17 |
| 91 | 25.00-25.49 | 25.24 | 0.15 | 43 | 11.6 | 212 | 28.5 | 46 | 17.2 | 11.1 | 1.63 |
| 100 | 25.50-26.03 | 25.79 | 0.16 | 44 | 12.1 | 210 | 25.9 | 48 | 19.4 | 11.1 | 1.28 |
| 103 | 26.04-26.49 | 26.27 | 0.14 | 42 | 12.1 | 203 | 28.5 | 41 | 16.9 | 11.4 | 1.59 |
| 107 | 26.50-26.94 | 26.72 | 0.13 | 41 | 11.3 | 204 | 28.4 | 40 | 19.0 | 11.2 | 1.32 |
| 107 | 26.95-27.34 | 27.15 | 0.12 | 40 | 13.9 | 203 | 28.4 | 43 | 18.8 | 11.5 | 1.58 |
| 106 | 27.35-27.70 | 27.52 | 0.11 | 40 | 10.4 | 204 | 23.5 | 40 | 16.6 | 11.5 | 1.47 |
| 106 | 27.71-28.19 | 27.97 | 0.14 | 39 | 11.6 | 198 | 24.7 | 37 | 15.3 | 11.6 | 1.29 |
| 107 | 28.20-28.60 | 28.39 | 0.12 | 39 | 11.4 | 196 | 23.1 | 38 | 17.9 | 11.7 | 1.37 |
| 106 | 28.61-29.00 | 28.80 | 0.11 | 39 | 12.3 | 199 | 26.6 | 36 | 16.6 | 11.5 | 1.25 |
| 107 | 29.01-29.46 | 29.22 | 0.12 | 40 | 10.4 | 202 | 22.5 | 38 | 16.3 | 11.6 | 1.50 |
| 107 | 29.47-29.94 | 29.72 | 0.14 | 39 | 12.6 | 196 | 27.9 | 37 | 17.8 | 11.6 | 1.45 |
| 107 | 29.95-30.39 | 30.15 | 0.13 | 37 | 11.7 | 192 | 25.0 | 34 | 14.9 | 11.9 | 1.41 |
| 106 | 30.40-30.81 | 30.59 | 0.13 | 38 | 10.0 | 192 | 23.7 | 32 | 13.8 | 12.0 | 1.33 |
| 101 | 30.82-31.27 | 31.03 | 0.14 | 35 | 11.6 | 187 | 26.8 | 31 | 14.8 | 12.2 | 1.50 |
| 98 | 31.28-31.78 | 31.49 | 0.14 | 33 | 11.8 | 178 | 26.9 | 30 | 12.1 | 12.4 | 1.65 |
| 95 | 31.80-32.68 | 32.25 | 0.26 | 35 | 10.5 | 183 | 27.8 | 31 | 13.6 | 12.0 | 1.44 |
| 87 | 32.69-33.79 | 33.24 | 0.32 | 33 | 11.6 | 184 | 29.1 | 25 | 13.4 | 12.6 | 1.79 |
| 68 | 33.81-35.18 | 34.46 | 0.39 | 29 | 11.1 | 177 | 22.5 | 26 | 13.3 | 12.5 | 1.30 |
| 54 | 33.52-39.87 | 36.62 | 1.27 | 29 | 12.3 | 177 | 26.5 | 24 | 11.5 | 12.7 | 1.55 |

† Abbreviations: N = the number of subjects in a lump; Mean = the lump mean values for the ranges; SD = standard deviation.

**TABLE 3** Lump mean values of the WHT2R ($\times 10^{-4}$ cm$^{-1}$), Sit-up (times/minute), SLJ (cm), 20-m MSSR (times), and 10-m SR (second) for females.

| N | WHT2R Range | WHT2R Mean | WHT2R SD | Sit-up Mean | Sit-up SD | SLJ Mean | SLJ SD | 20-m MSSR Mean | 20-m MSSR SD | 10-m SR Mean | 10-m SR SD |
|---|---|---|---|---|---|---|---|---|---|---|---|
| 55 | 21.06-24.18 | 23.29 | 0.78 | 30 | 12.4 | 153 | 29.3 | 25 | 12.9 | 13 | 1.45 |
| 64 | 24.22-25.09 | 24.66 | 0.27 | 29 | 11.1 | 154 | 21.7 | 26 | 12.7 | 13.1 | 1.49 |
| 85 | 25.11-25.89 | 25.49 | 0.23 | 31 | 12.3 | 156 | 26.7 | 27 | 11 | 13 | 1.34 |
| 90 | 25.90-26.55 | 26.23 | 0.18 | 30 | 10.8 | 149 | 21.7 | 24 | 10.8 | 13.3 | 1.35 |
| 94 | 26.56-27.14 | 26.84 | 0.17 | 29 | 10.6 | 150 | 22.3 | 24 | 11.1 | 13.5 | 1.43 |
| 94 | 27.16-27.64 | 27.39 | 0.15 | 31 | 9.76 | 150 | 20.3 | 24 | 11.3 | 13.5 | 1.43 |
| 96 | 27.65-28.15 | 27.88 | 0.15 | 26 | 12 | 147 | 23.8 | 22 | 10.3 | 13.7 | 1.77 |
| 95 | 28.16-28.64 | 28.4 | 0.16 | 26 | 12.1 | 146 | 26.5 | 23 | 10.7 | 13.5 | 1.62 |
| 96 | 28.65-29.07 | 28.87 | 0.13 | 27 | 10.7 | 146 | 28.4 | 21 | 9.74 | 13.9 | 1.74 |
| 95 | 29.09-29.47 | 29.29 | 0.11 | 27 | 11.5 | 145 | 23.9 | 22 | 10.8 | 13.6 | 1.39 |
| 94 | 29.48-30.02 | 29.75 | 0.16 | 26 | 12.4 | 144 | 27 | 22 | 12.2 | 13.7 | 1.49 |
| 95 | 30.03-30.45 | 30.24 | 0.14 | 24 | 10.9 | 143 | 22.3 | 21 | 10.1 | 13.9 | 1.59 |
| 96 | 30.46-30.94 | 30.72 | 0.14 | 24 | 10.9 | 139 | 24 | 20 | 11.5 | 14.1 | 1.85 |
| 94 | 30.96-31.48 | 31.21 | 0.15 | 26 | 10.6 | 142 | 20.4 | 20 | 9.35 | 13.8 | 1.3 |
| 96 | 31.49-31.96 | 31.73 | 0.14 | 27 | 10.7 | 141 | 23.2 | 19 | 9.04 | 14.1 | 1.65 |
| 94 | 31.97-32.51 | 32.25 | 0.16 | 23 | 10.7 | 138 | 22.5 | 19 | 9.46 | 14.3 | 1.61 |
| 94 | 32.54-33.05 | 32.79 | 0.15 | 21 | 10.7 | 134 | 24.5 | 17 | 8.04 | 14.4 | 1.51 |
| 94 | 33.07-33.67 | 33.37 | 0.18 | 23 | 12.2 | 140 | 23.5 | 18 | 11.3 | 14 | 1.58 |
| 90 | 33.68-34.53 | 34.08 | 0.25 | 20 | 10.8 | 131 | 25.9 | 16 | 10.4 | 14.7 | 1.73 |
| 92 | 34.54-35.54 | 34.96 | 0.27 | 21 | 9.95 | 134 | 23.8 | 16 | 6.95 | 14.5 | 1.81 |
| 88 | 35.56-36.79 | 36.17 | 0.37 | 17 | 9.17 | 125 | 24.1 | 15 | 6.79 | 14.8 | 1.64 |
| 80 | 36.81-38.38 | 37.6 | 0.49 | 19 | 10.7 | 129 | 22.4 | 13 | 5.62 | 14.8 | 1.65 |
| 67 | 38.42-43.58 | 40.03 | 1.24 | 15 | 8.21 | 118 | 25 | 13 | 7.1 | 15.6 | 1.97 |

† Abbreviations: N = the number of subjects in a lump; Mean = the lump mean values for the ranges; SD = standard deviation.

The correlation coefficients between the indices and fitness tests were calculated using correlation analyses in Microsoft Excel 2014, and linear regression was carried out using Sigmaplot 14. The distribution of fitness with respect to WHT2R is quite complicated. Therefore, the lump mean values (tables 2 and 3) were considered to investigate the dependence of fitness on the WHT2R. Male and female subjects were divided into groups containing approximately 105 and 95 subjects with successive WHT2Rs except near the two endpoints. The number of subjects in each group was slightly different because the number of subjects with the same WHT2R is not uniform and decreases toward the two endpoints of the WHT2R because of a lack of subjects.

# 3 RESULTS

## 3.1 Correlation between the indices and fitness tests

Figure 1 shows the correlation coefficients ($r$) between the indices and fitness tests. A more significant correlation coefficient indicates a stronger correlation between the two values. Therefore, figure 1 shows that the WHT2R has a stronger correlation with all the fitness tests than the other four indices. Sit-ups were negatively correlated with all five indices for both males and females. This means that the number of sit-ups performed in the test decreases with increasing index. The BMI increases with increasing weight, while ABSI, WHT.5R, WHtR, and WHT2R increase with increasing waist circumference, and the muscular endurance decreases with decreasing number of sit-ups. Therefore, muscular endurance decreases with increasing weight or waist circumference.

The SLJ performance was negatively correlated with all five indices for both males and females. This means that the SLJ score decreases with increasing index. Therefore, the speed decreases with increasing weight or waist circumference because it decreases with decreasing SLJ score. The 20-m MSSR performance was also negatively correlated with all five indices for males and females. This means that the 20-m MSSR performance decreases with increasing index. Therefore, cardiorespiratory endurance decreases with increasing weight or waist circumference because the cardiorespiratory endurance decreases with decreasing 20-m MSSR performance. In contrast, the 10-m SR time was positively correlated with all five indices for both males and females. This means that the 10-m SR time increases with increasing index. Hence, the speed and agility decrease with increasing weight or waist circumference because an increase in the 10-m SR time means a decrease in speed and agility.

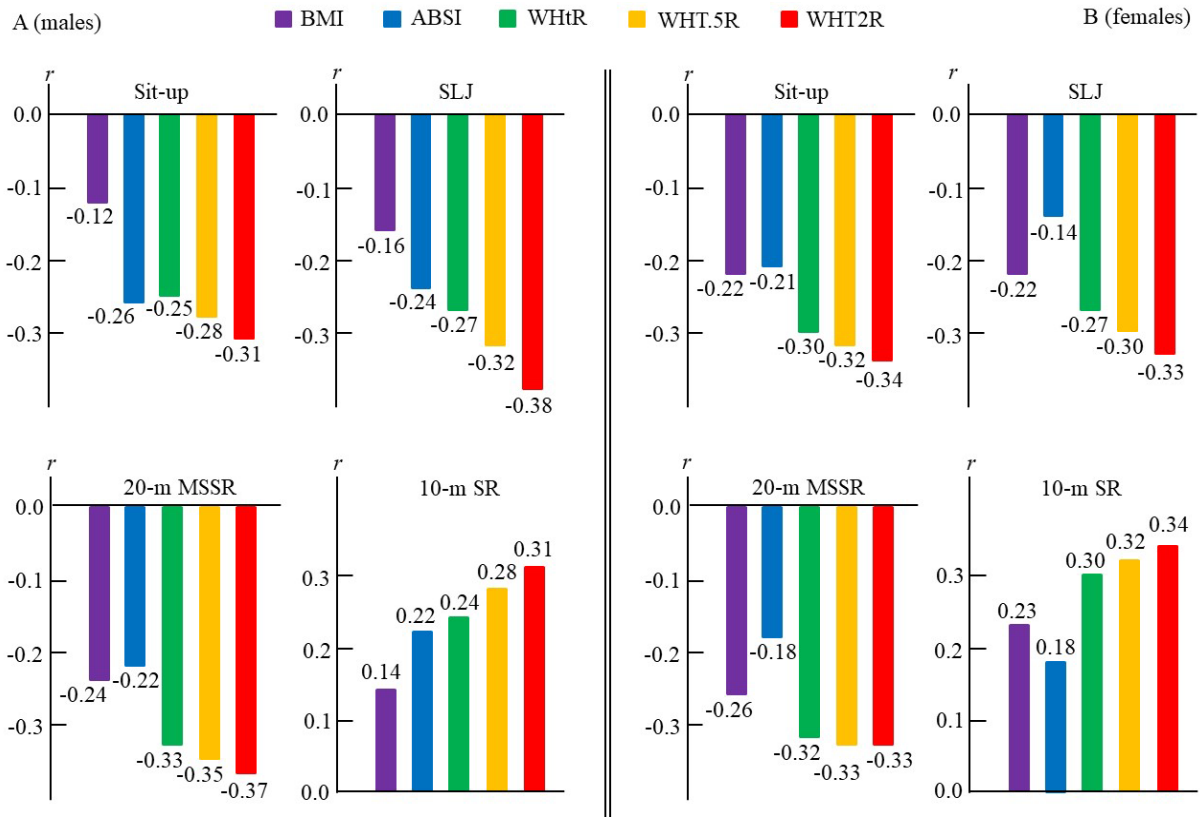

**FIGURE 1**

A. Correlation coefficients ($r$) between the indices and fitness tests for males.

B. Correlation coefficients ($r$) between the indices and fitness tests for females.

### 3.2 Correlation between the WHT2R and fitness tests per generation

Figure 2 shows the correlation coefficients ($r$) between the WHT2Rs and fitness tests per generation. The correlations for males were highest in those in their 30s in all the fitness tests and then decreased with age. The correlations for females increased with age and were highest in those in their 50s in all fitness tests.

In Korea, every healthy male youth must perform military service, where they exercise regularly. Therefore, the increase in abdominal obesity and the decrease in fitness by the change in life pattern after military service, as well as aging, can explain the highest correlations for males in their 30s. The decreases in the correlations in males older than 40 years can be interpreted by the increase in the concern for health and aging. This means that the correlations increase as males age due to the increase in abdominal obesity and the decrease in fitness, but these increases are curtailed by the decrease in abdominal obesity and increase in fitness due to physical activity or exercise (figure 2A).

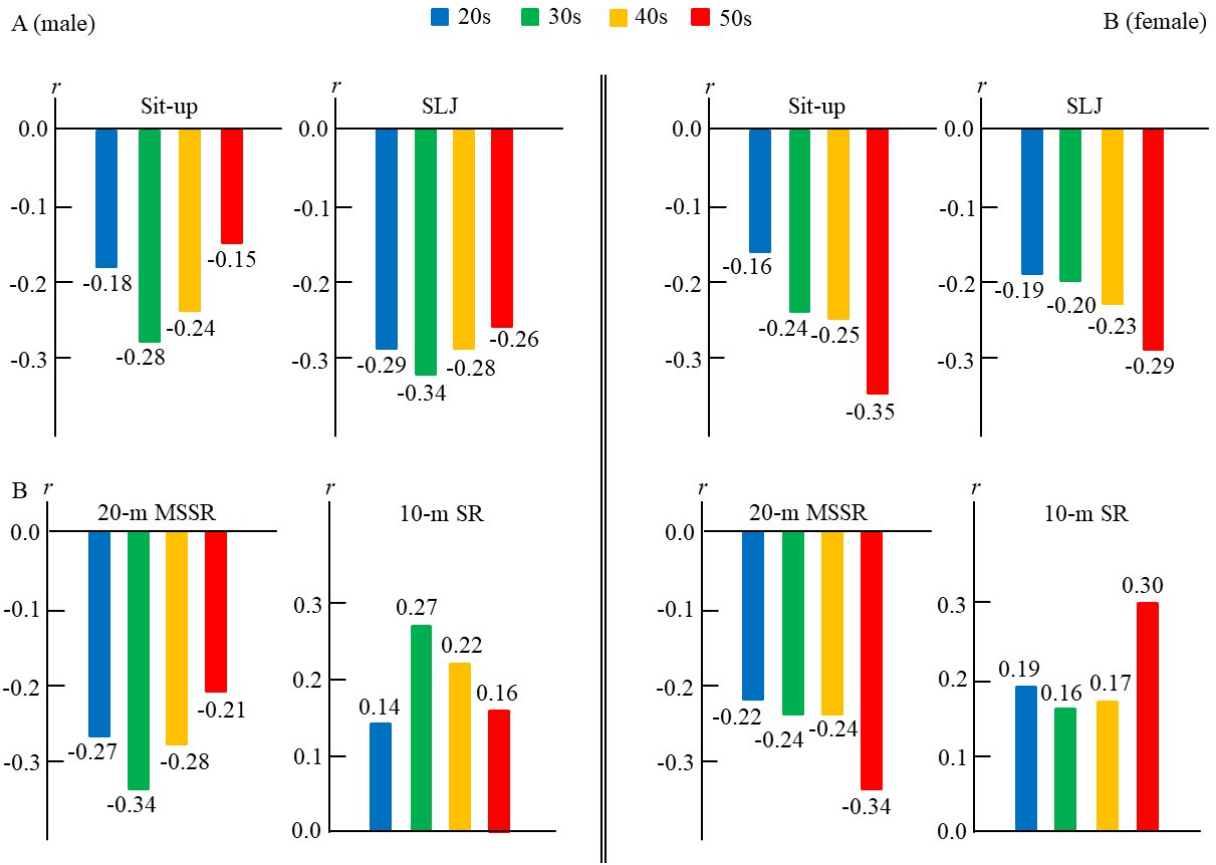

**FIGURE 2**

A. Correlation coefficients between the WHT2R and fitness tests for males per generation.

B. Correlation coefficients between the WHT2R and fitness tests for females per generation.

The reasons why the correlations increase with age for females can be attributed to the increase in age for childbirth and concerns for body shape as well as aging. The correlations for females are relatively small in their 20s - 40s owing to the distorted association between fitness and body shape occurred because they reduced their abdominal obesity by controlling food intake for a good body shape but they did not increase their fitness by exercise. On the other hand, the correlation increases from 40 years due to the increase in abdominal obesity and the decrease in fitness after childbirth (figure 2B).

## 3.3 Analysis using the lump mean values

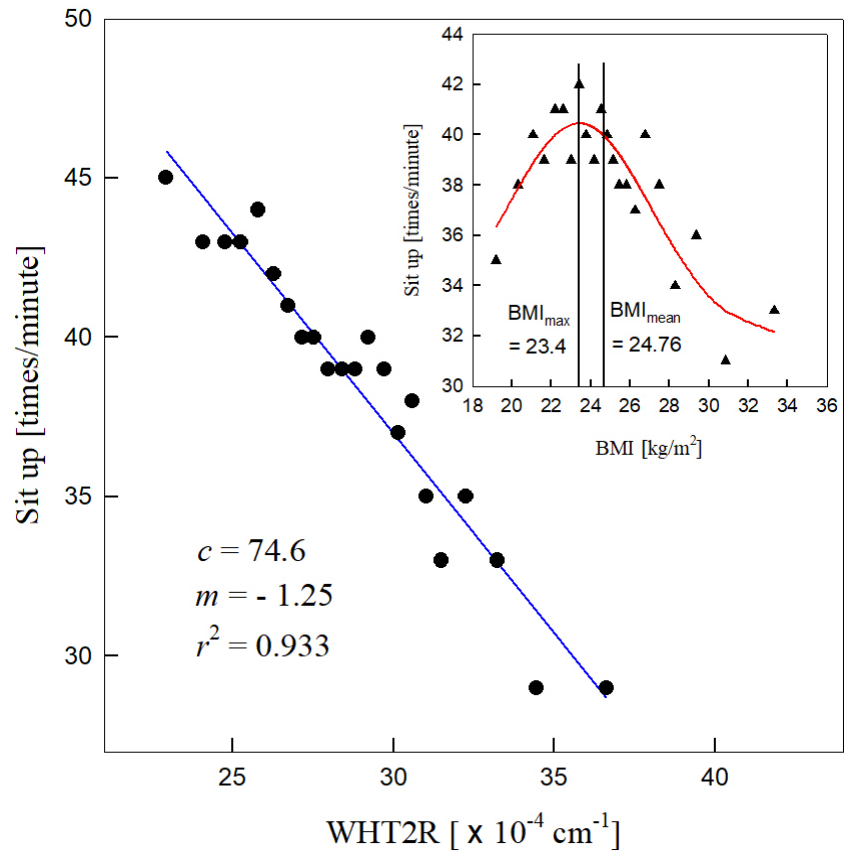

**FIGURE 3** Black circles are the LMVs of the sit-up test for the LMVs of the WHT2R for males, and the solid blue line denotes the straight line for the black circles fitted to Eq. (2). The inset shows a previous result for the BMI (Hwang et al., 2020).

The dependence of fitness on WHT2R was investigated using the lump mean values (LMVs) (tables 2 and 3). Figure 3 shows that the LMV of the sit-up test decreases linearly with increasing LMVs of WHT2R among males. On the other hand, the LMVs of the sit-up test increased as the LMVs of BMI among males increased to $BMI_{max}$ and then decreased as the BMI increased from $BMI_{max}$ showing a Gaussian distribution, where $BMI_{max}$ (= 23.4 × $10^{-4}$ cm$^{-1}$) denotes the BMI for the best fitness (Hwang et al., 2020) and it was smaller than the mean value of BMI ($BMI_{mean}$ = 24.76 × $10^{-4}$ cm$^{-1}$). The condition for the best fitness could be obtained using the LMV of the BMI because the LMVs of the sit-up test with respect to the LMVs of the BMI was fitted to a Gaussian curve, as shown in the inset, while the LMVs of the sit-up test with respect to the LMVs of WHT2R can be fitted to a straight line, as shown in figure 3, which is expressed as:

$$y(x) = mx + c \tag{2}$$

where $x$, $c$, $m$, and $y(x)$ are the LMV of WHT2R, a constant, slope, and LMVs of the four fitness tests. Variation of the LMVs of fitness tests with respect to that of the LMVs of WHT2R increases with increasing $m$. $m$ is positive for a positive correlation and negative for a negative correlation. The figure shows the two parameters defined in Eq. (2), $m$ and $c$. The values at the two end points are unstable because of a lack of subjects. The discrepancies between the black circles and solid blue line can be reduced if a sufficiently large number of subjects are available. Figure 3 shows that the goodness-of-fit for linear regression is sufficient because the coefficients of determination ($r^2$) are sufficiently large.

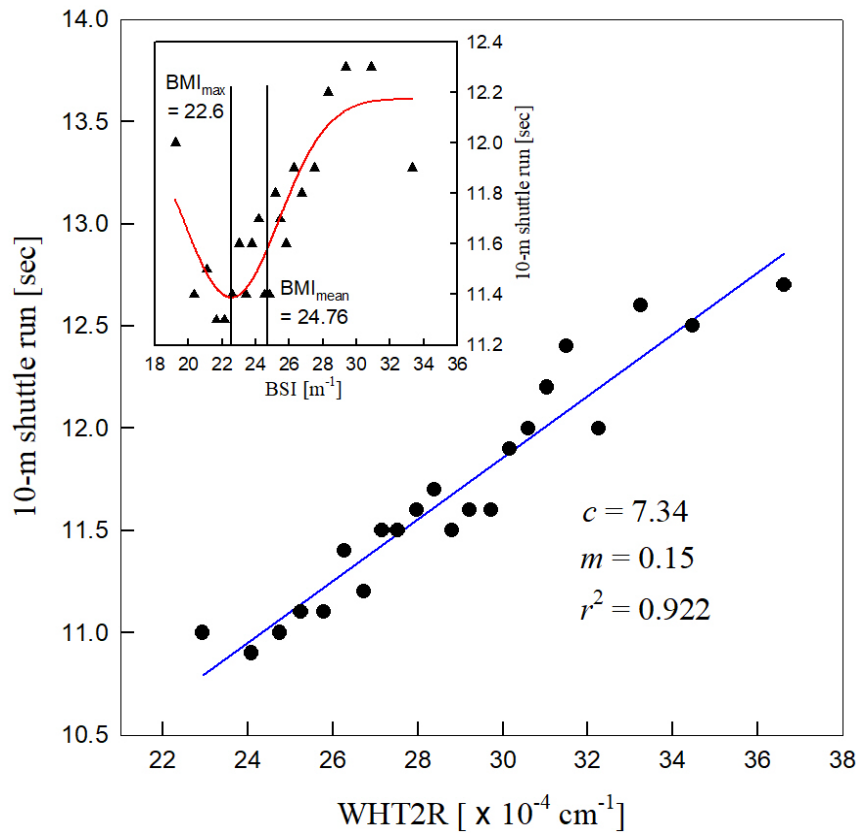

**FIGURE 4** Black circles are the lump mean values of the 10-m SR test for the lump mean WHT2R for males, and the solid blue line denotes the straight line for the black circles fitted to Eq. (2). The inset shows the result for BMI (Hwang et al., 2020).

The LMVs of the 10-m SR test with respect to the LMVs of WHT2R among males can be fitted to a straight line, as shown in figure 4, whereas the LMV of the 10-m SR test decreased as the LMVs of the BMI among males increases to $BMI_{max}$ and then increased with further increases in BMI, as in an inverse Gaussian distribution as shown in the inset. The LMVs of SLJ and 20-m MSSR tests with respect to WHT2R can be fitted to the straight lines in the same manner. Table 4 lists the results, including those for women. The results show that the LMVs of all fitness tests can be fitted better to a straight line than a Gaussian curve using the LMV of the WHT2R rather than using the LMV of the BMI. This means that setting the goal of exercise using the WHT2R is easier and more convenient than using the BMI. The smallest WHT2R is the WHT2R for the best fitness. On the other hand, WHT2R is not applicable to establish an exercise aim of people with very weak fitness because it does not include the muscle, whereas the BMI is applicable because it includes the muscle and fat in the weight (Hwang et al., 2020) if other effects, such as the lung capacity, affecting the fitness score are neglected.

### 3.4 Analysis using the fitness sensitivity percentage to WHT2R

Although the goodness-of-fit for linear regressions are sufficient by the coefficients of determination (table 4), the regression equation [Eq. (2)] cannot be used to establish the goal of exercise. In order to establish the goal of exercise, fitness sensitivity percentage to WHT2R (FSPW) is introduced:

$$\text{FSPW (WHT2R}_{sub}) \equiv \left| \frac{y_{subW} - y_{dW}}{y_{subW}} \right| \times 100\ [\%] \tag{3}$$

where $y_{subW}$ is the fitness test score corresponding to the WHT2R of each subject (WHT2R$_{sub}$) and $y_{dW}$ is the score corresponding to the desired value of the WHT2R, which was chosen as the minimum of the WHT2R (WHT2R$_{min}$) for convenience in this study, but it can be chosen differently according to the desired value of each subject. As shown in figure 3, WHT2R$_{min}$ is the WHT2R when the fitness is best and corresponds to the $BMI_{max}$. FSPW means the ratio of each fitness test score to increase (sit-up, SLJ, 20-m MSSR) or decrease (10-m SR) in order to reach the WHT2R$_{min}$. Table 5 lists the results for WHT2R$_{sub}$ =

30 and 32 ($\times 10^{-4}$ cm$^{-1}$) (males) and WHT2R$_{sub}$ = 31 and 34 ($\times 10^{-4}$ cm$^{-1}$) (females). For example, a male with WHT2R$_{sub}$ = 30 $\times 10^{-4}$ cm$^{-1}$ needs to increase 23.7% of his sit-up score to attain WHT2R$_{min}$ = 22.93 $\times 10^{-4}$ cm$^{-1}$ and a male with WHT2R$_{sub}$ = 32 $\times 10^{-4}$ cm$^{-1}$ needs to increase 32.7% of his sit-up score to attain the same WHT2R$_{min}$.

**TABLE 4** $m$ and $c$ for fitting the lump mean values of the fitness tests ($y$) with respect to the lump mean values of the WHT2R ($x$) to a straight line, $y(x) = mx + c$.

|  | Male | | | | Female | | | |
|---|---|---|---|---|---|---|---|---|
|  | Sit-up | SLJ | 20-m MSSR | 10-m SR | Sit-up | SLJ | 20-m MSSR | 10-m SR |
| $m$ | -1.25 | -3.5 | -2.09 | 0.15 | -0.99 | -2.19 | -0.92 | 0.15 |
| $c$ | 74.6 | 298.4 | 97.4 | 7.34 | 55.1 | 208.5 | 48.4 | 9.41 |
| $r^2$ | 0.933 | 0.922 | 0.933 | 0.922 | 0.883 | 0.937 | 0.949 | 0.937 |

Note: The dimension of $c$ is same as that of each fitness test, and the dimension of $m$ is the same as the dimension of each fitness test divided by cm. $r^2$ = coefficients of determination.

**TABLE 5** Fitness sensitivity percentage to WHT2R for each fitness test.

|  | Male | | | | Female | | | |
|---|---|---|---|---|---|---|---|---|
|  | Sit-up | SLJ | 20-m MSSR | 10-m SR | Sit-up | SLJ | 20-m MSSR | 10-m SR |
| $y_{dW}$ | 45.9 | 218.1 | 49.5 | 10.8 | 32 | 157.5 | 27 | 12.9 |
| $y_{subW1}$ | 37.1 | 193.4 | 34.7 | 11.8 | 24.4 | 140.6 | 19.9 | 14.1 |
| $y_{subW2}$ | 34.6 | 186.4 | 30.5 | 12.1 | 21.4 | 134 | 17.1 | 14.5 |
| FSPW$_1$ [%] | 23.7 | 12.8 | 42.7 | 9.3 | 31.1 | 12 | 35.7 | 9.3 |
| FSPW$_2$ [%] | 32.7 | 17 | 62.3 | 12 | 49.5 | 17.5 | 57.9 | 12.4 |

† Abbreviations: $y_{dW}$ = fitness test score corresponding to the desired value of WHT2R, and $y_{subW}$ = fitness test score corresponding to the WHT2R of each subject.

Note: The subscript 1 denotes the results for WHT2R$_{sub1}$ = 30 (males) and 31 (females) ($\times 10^{-4}$ cm$^{-1}$) and the subscript 2 denotes the results for WHT2R$_{sub2}$ = 32 (male) and 33 (females) ($\times 10^{-4}$ cm$^{-1}$). The dimensions of $y_{dW}$, $y_{subW1}$, and $y_{subW2}$ are same as the dimension of each fitness test. [Units: sit-up (times/minute), SLJ (cm), 20-m MSSR (times), 10-m SR (second)]

The sensitivity of WHT2R to fitness increases with decreasing FSPW, i.e., the WHT2R can be reduced more by decreasing the same ratio of the fitness score because the FSPW is smaller.

Therefore, FSPW corresponds to the $b$ value in BMI analysis, which is the width of the Gaussian curve and means the sensitivity of the fitness to the BMI (Hwang et al., 2020). Table 5 shows that the WHT2R can be controlled effectively by decreasing the 10-m SR score, e.g., a 9.3% decrease in the 10-m SR score for males with $\text{WHT2R}_{\text{sub}} = 30 \times 10^{-4}$ cm$^{-1}$ and 9.3% decrease in the for females with $\text{WHT2R}_{\text{sub}} = 31 \times 10^{-4}$ cm$^{-1}$ are necessary to attain $\text{WHT2R}_{\text{min}} = 22.93 \times 10^{-4}$ cm$^{-1}$(males) and $\text{WHT2R}_{\text{min}} = 23.29 \times 10^{-4}$ cm$^{-1}$(females), respectively. On the other hand, it does not mean that reducing the 10-m SR score is the easiest way to attain the desired WHT2R values kinematically; it just means that the desired values of WHT2R can be achieved effectively by decreasing the 10-m SR score. Increasing the ratios of the 20-m MSSR score is necessary to attain the same desired values of the WHT2R, e.g., a 42.7% increase for males with $\text{WHT2R}_{\text{sub1}} = 30 \times 10^{-4}$ cm$^{-1}$ and 35.7% increase for females with $\text{WHT2R}_{\text{sub2}} = 31 \times 10^{-4}$ cm$^{-1}$ are needed to achieve the desired values of the WHT2R.

## 4 DISCUSSION

Overall, the sit-up, SLJ, and 20-m MSSR performance decreased while the 10-m SR increased as the BMI, ABSI, WHT.5R, WHtR, and WHT2R were increased. Therefore, the sit-up, SLJ, and 20-m MSSR performance were negatively correlated with the indices, whereas the 10-m SR was positively correlated with them. The correlation coefficients between the BMI and the four fitness tests for females were larger than those for males. The four fitness tests were the most weakly correlated with the BMI for males and the ABSI for females. On the other hand, the WHT.5R, WHtR, and WHT2R were more associated with all the fitness tests than the BMI and ABSI, and there was a stronger correlation between all the fitness tests and the WHT2R than the WHT.5R and WHtR for both males and females. Therefore, the WHT2R is a more accurate index for assessing fitness than the BMI, ABSI, WHT.5R, and WHtR.

The waist-to-height ratio (WC/height$^b$) becomes more associated with all the fitness tests as the scaling exponent ($b$) increases (figure 1) but the WHT3R, which is defined as the WC divided by the cube of the height ($b = 3$), did not show a stronger association with all of the fitness tests than the WHT2R. The correlation coefficients ($r$) between the WHT3R and fitness tests were $r = -0.31$ (sit-up), $-0.40$ (SLJ), $-0.35$ (20-m MSSR), and $0.31$ (10-m SR) for males, and $r = -0.33$ (sit-up), $-0.34$ (SLJ), $-0.32$ (20-m MSSR), and $0.34$ (10-m SR) for females. Therefore, the waist-to-height ratio (WC/height$^b$) with $b$ values between 2 and 3 will have the best association with all of the fitness tests. This needs to be studied more using the allometric model (Bustamante et al., 2015) to identify the optimal body size and shape.

Although the WHT2R has the best association with all of the fitness tests, it does not mean that it has an acceptable goodness of fit because the coefficient of determination ($r^2$) is quite low ($r^2 \sim 0.1$). On the other hand, the associations for other four indices (BMI, ABSI, WHT.5R, WHtR) were smaller than that of the WHT2R. Therefore, the WHT2R can contribute more effectively to explaining fitness and obesity than the other indices if additional factors, such as sex, smoking, diabetes, blood pressure, and other factors are considered and if more advanced statistical methods are used. Although the reasons for the strong correlations for males in their 30s and females in their 50s were attributed to military service for males and childbirth for females, more study considering aging, cardiopulmonary function, morphological characteristic, and muscular strength will be needed.

The correlation of the LMV of the fitness with that of the WHT2R was fitted to a simple straight line in the present study, whereas that with the LMV of the BMI was fitted to a Gaussian curve in the previous study (Hwang et al., 2020). Therefore, the correlation with fitness was interpreted more simply using the LMV of the WHT2R rather than that of the BMI. The desired WHT2R value could be obtained simply and uniquely from the minimum values of a straight line corresponding to the best values of the sit-up, SLJ, 20-m MSSR, and 10-m SR test scores, while they were different according to the best values of the each fitness test given by the Gaussian curves (Hwang et al., 2020). The ratio of the fitness test score to increase (sit-up, SLJ, 20-m MSSR) or decrease (10-m SR) in order to attain the goal of exercise could be obtained using the FSPW. Although a small FSPW means that the desired WHT2R value can be achieved effectively with a decreasing fitness test score, it does not mean that the fitness test with the smallest FSPW is the easiest exercise to reduce the WHT2R kinematically; it just means that the ratio of the fitness test score to increase (sit-up, SLJ, 20-m MSSR) or decrease (10-m SR) to achieve the desired WHT2R value is small, but it can be used as a guide for the goal of exercise.

# 5 CONCLUSIONS

This study showed that waist circumference rather than body weight should be used in a body index to assess fitness. Furthermore, this study showed that the waist circumference is more appropriate as an index when it is divided by the square of the height than when it is divided by the height or square root of the height. A new waist-to-height ratio (WHT2R) introduced in this study was a more effective index for assessing fitness than the body mass index and other waist-to-height ratios. Setting the goal of exercise using the WHT2R was easier and more convenient than using the BMI because the fitness with respect to the WHT2R could be fitted to a simple straight line while the fitness with respect to the BMI was fitted to the Gaussian curve. On the other hand, the BMI has merit in that it would be applicable to establish the exercise goal of people with poor fitness. Therefore, it is expected that the WHT2R could be used as a more effective ratio to determine the fitness and establish an exercise aim if the BMI is used in a complementary manner. The WHT2R, LMV, and FSPW may be used as a useful indicator for fitness and obesity of other races with similar physiques to set their exercise aims.

# ACKNOWLEDGEMENTS

The author would like to thank the Korea Sports Promotion Foundation for providing the data from the 2017 survey of national physical fitness. This research did not receive any specific grants from funding agencies in the public, commercial, or non-profit sectors.

# FUNDING

None.

# CONFLICTS OF INTEREST

None.